\documentclass[10pt,a4paper]{article}

\usepackage{amssymb}			
\usepackage{amsmath}			
\usepackage{graphicx}			
\usepackage[latin1]{inputenc}	
\usepackage{hyperref}			
\usepackage[usenames,dvipsnames,svgnames,table]{xcolor}						
\hypersetup{colorlinks=true, citecolor=Blue, linkcolor=Blue, urlcolor=Blue}				

\usepackage{graphicx}
\usepackage{color}

\usepackage{amsmath,amsfonts,amssymb,amsthm,epsfig,epstopdf,url,array}
\usepackage{float}

\usepackage{wrapfig}
\usepackage{cancel}

\newtheorem{thm}{Theorem}[section]

\newtheorem{defn}{Definition}[section]

\newtheorem{prop}{Proposition}[section]

\newcommand{\R}{{\mathbb R}}
\newcommand{\q}{{\quad}}

\newcommand{\be}{\begin{equation}}
	\newcommand{\ee}{\end{equation}}

\def\be{\begin{equation}}
	\def\ee{\end{equation}}

\newtheorem{corollary}{Corollary}[section]
\newcommand{\RomanNumeralCaps}[1]

\begin{document}
	
	\title{Nonlinear monotone energy stability of plane shear flows: Joseph or Orr critical thresholds?}
	\author{Giuseppe Mulone\footnote{Universit\`{a} degli Studi di Catania (retired), Dipartimento di Matematica e Informatica, Viale Andrea Doria 6, 95125 Catania, Italy,  giuseppe.mulone@unict.it}}
	\date{}
	\maketitle

%
%
%

\begin{abstract}
Critical  Reynolds numbers for the monotone exponential energy stability of Couette and Poiseuille plane flows were obtained by Orr 1907 \cite{Orr1907} in a famous paper, and by Joseph 1966 \cite{Joseph1966},  Joseph and Carmi 1969 \cite{JosephCarmi1969} and Busse 1972 \cite{Busse1972}. All these authors obtained their results applying variational methods to compute the maximum of a functional ratio derived  from the Reynolds-Orr energy identity.
Orr and Joseph obtained different results, for instance in the Couette case Orr computed the critical Reynolds value of  44.3 (on spanwise perturbations) and Joseph 20.65 (on streamwise perturbations).

Recently in Falsaperla et al. 2022 \cite{FalsaperlaMulonePerrone2022}, the authors conjectured that the search of the maximum should be restricted to  a subspace of the space of kinematically admissible perturbations. With this conjecture, the critical nonlinear energy  Reynolds number was found among spanwise perturbations (a Squire theorem for nonlinear systems).

With a direct proof and an appropriate and original decomposition of the dissipation terms in the Reynolds-Orr identity we show the validity of  this conjecture  in the space of three dimensional perturbations.     
\end{abstract}

\parindent 0pt



\section{Introduction}
\label{sec:intro}

The study of the stability of shear flows is important in many applications and it has been object of study over the past 150 years, see e.g.,  \cite{Joseph1976}, \cite{DrazinReid2004}, \cite{SchmidHenningson2001}, \cite{Eckhardt.et.al2007}.

The laminar flows of Couette and Poiseuille between parallel planes are of particular importance. Their stability and instability have been extensively studied in both the linear and non-linear cases \cite{Romanov1973}, \cite{Orszag1971}, \cite{Orr1907}, \cite{Joseph1966}, \cite{JosephCarmi1969}, \cite{Busse1972}. See also the recent papers \cite{Fuentes.Goluskin.Chenishenko.2022}, \cite{Nagy.2022}, \cite{Xiong.Chen.2019}.

The critical linear Reynolds numbers for the onset of instability (obtained with the spectral-modal method) and critical non-linear values (energy method) are very different from one another and are very different from the critical values of the experiments. This is often indicated as subcritical instability.  A possible mechanism to explain the subcritical transition in Couette and Poiseuille flows is the transient growth of the  energy of the perturbation velocity field. Recently the transient regime has been extensively studied, see e.g., \cite{ButlerFarrel1992}, \cite{ReddyHenningson1993}, \cite{SchmidHenningson2001}, \cite{Eckhardt.et.al2007}, \cite{Martinelli2011}, \cite{Tuckerman.et.al2020}, \cite{Klotz.et.al2021}.
It is therefore important to \textit{determine the exact value} of the critical energy Reynolds  number $ {\rm Re}_E $ which is the largest Reynolds number under which the energy is monotonically decreasing. Below such critical value the energy  cannot display transient growth.

The critical Reynolds numbers  $ {\rm Re}_E $ for monotone energy stability of plane Couette and Poiseuille flows have been obtained by Orr \cite{Orr1907} in a famous paper published in 1907, and by Joseph \cite{Joseph1966},  Joseph and Carmi \cite{JosephCarmi1969} and  Busse \cite{Busse1972} in the late sixties with variational methods applied to a functional ratio which comes from the Reynolds-Orr energy identity.
These authors obtained different results.
\newpage

In his paper \cite{Orr1907}, Orr writes: ``\textit{Analogy with other problems leads us to assume that disturbances in two dimensions will be less stable than those in three; this view is confirmed by the corresponding result in case viscosity is neglected}". He also says:  ``\textit{The three-dimensioned case was attempted, but it proved too difficult}".
Orr considers two-dimensional spanwise perturbations: $v\equiv 0$ and ${\partial}_y\equiv 0$ (see also \cite{Squire1933},  \cite{DrazinReid2004}).
He finds the critical value  Re$^{x}=44.3$ in the Couette case:  \textit{the critical Reynolds number with respect to spanwise perturbations}, see  \cite[p. 128]{Orr1907}, \cite[p. 181]{Joseph1976}.

In his monograph \cite[p. 181]{Joseph1976}, Joseph says: ``\textit{Orr's assumption about the form of the disturbance which increases at the smallest {\rm Re} is not correct since we shall see that the energy of an $x$-independent disturbance (streamwise perturbations) can increase when  ${\rm Re} >2 \sqrt{1708}\simeq 82.65$}" (in our dimensionless form Re$^y$=20.6)

Busse \cite[p. 29]{Busse1972} in his paper writes 
``\textit{Numerical computations suggest that the eigenvalue $R_E$ is attained for $x$-independent solutions. Since this result has contradicted the physical intuition of earlier investigators in this field, it is desirable to find a rigorous proof for this property}". However he remarks in a note on p. 29:  ``\textit{Joseph first found that the minimizing solution in the case of Couette flow was independent of $x$. He also gave a proof of this fact. A gap of his proof has been found, however, recently by J. Serrin (private communication by D. D. Joseph) who pointed out that Joseph did not account for the possibility that the required minimum could appear at the end point $k_x=k$}".  

Drazin and Reid in their monograph,  \cite[pp. 429--430]{DrazinReid2004}, write:
``\textit{Orr (1907) further assumed that two-dimensional perturbations, i.e. the least eigenvalue of equations (53.21) can be obtained as the least eigenvalue of equation (53.22). Although Squire's theorem provides a justification for this assumption in the case of linear theory, it is not applicable to equations (53.21) and the assumption is false in general. 
	\\
	The determination of the least eigenvalue of equations (53.21) is clearly a formidable problem in general and results are known for only a few flows. In the case of Couette flow with $U(z)=z$  $(-1\le z\le 1)$, however, Joseph (1966) has proved that the least eigenvalue of equations (53.21) is still associated with a two-dimensional disturbance but one which varies only in the $yz$-plane, i.e. the perturbed flow consists of rolls whose axes are in the direction of the basic flow."}

Schmid and Henningson, \cite[p. 192]{SchmidHenningson2001}, write:
``\textit{Joseph (1976) showed that the largest Reynolds numbers ${\rm Re}_E$ result for disturbances that are independent of the streamwise direction."}

The critical energy Reynolds number is obtained by determining the  maximum of a functional ratio obtained from the Reynolds-Orr identity. Orr assumed (without proving it) that the maximum is obtained on the perturbations to the velocity field which satisfy no-slip boundary conditions,  are periodic, divergence-free, and they are spanwise. Joseph instead assumed that the maximum is attained on streamwise perturbations.

This problem was recently taken up by \cite{FalsaperlaGiacobbeMulone2019} and  by \cite{FalsaperlaMulonePerrone2022}. 
In particular, in \cite{FalsaperlaMulonePerrone2022}, the authors  conjectured that the search  for the maximum should be performed on a subspace of the space of kinematically admissible perturbations: the space of physically admissible functions competing for the maximum. With this conjecture the critical energy  Reynolds number was found among the two-dimensional spanwise perturbations as Orr assumed.

We note that Lorentz \cite{Lorentz.1907}  made these observations, also reported in Lamb \cite{Lamb.1924} , p. 640: ``One or two consequences of the energy equation may be noted. In the first place, the relative magnitude of the two terms on the right-hand side  is unaffected if we reverse the signs of $u, v, w$, or if we multiply them by any constant factor. The stability of a given state of mean motion should not therefore depend on the \textit{scale} of the disturbance. On the other hand, certain combinations of  $u, v, w$, appear to be more favourable to stability than others". Therefore, in the study of the following maximum problems we will always assume that this \textit{scale invariance} property holds.

It should be noted here that in the Joseph's calculations the maximizing functions found \textit{do not verify} this property of invariance of scale. In fact, $u$ can be replaced by $-u$ (because in his maximum problem $u_x=0$) leaving the same $w$.

In this article we  prove  the conjecture of \cite{FalsaperlaMulonePerrone2022} is true in  two simple mathematical ways (see Section 3).  

First, we give a simple proof by showing that the maximum is obtained when the second component $v$ of the perturbation to the velocity field  is zero and the maximizing perturbation is spanwise.

Second, the main idea is  to partition the dissipative terms: we split the dissipation term $$-{\rm Re}^{-1}\Vert \nabla {\bf u}\Vert^2$$ in the Reynolds-Orr identity into two terms: $- {\rm Re}^{-1} [\Vert \nabla u \Vert^2+\Vert \nabla w \Vert^2]$ and $ -{\rm Re}^{-1}\Vert \nabla v \Vert^2$ and to study an appropriate maximum of a functional ratio on the space of kinematically admissible perturbations, see maximum problem \eqref{maxRefin}. 

It follows that the Orr's results are correct, the critical Reynolds numbers for monotone energy stability are ${\rm Re}_{Orr}=44.3$ (Couette flow) and  ${\rm Re}_{Orr}=87.6$ (Poiseuille flow). 
 Other than being a mathematical argument, this fact can be compared with the instability thresholds obtained experimentally in \cite{Prigent.et.al2003} for tilted perturbations (see Section 4).

Moreover, a \textit{Squire theorem holds for nonlinear monotone energy stability}: the least stabilizing perturbations are the two-dimensional spanwise perturbations.

Our result is compatible with and reinforces the recent result by Fuentes et al. \cite{Fuentes.Goluskin.Chenishenko.2022}.

The plan of the paper is as follows. 
In Sec. 2 we write the non-dimensional perturbation equations of laminar flows between two horizontal planes with no-slip boundary conditions, and we recall the classical linear stability/instability results.
In Sec. 3 we study monotone nonlinear energy stability and prove that the correct critical energy Reynolds number are those obtained by Orr on the spanwise perturbations. As a consequence a Squire Theorem holds in the nonlinear case. In Sec. 4  we  make some final comments.

\section{Laminar flows between two parallel planes}
Given  a reference frame $Oxyz$, with unit vectors ${\bf i},{\bf j}, {\bf k}$, consider the layer $\mathcal D = \R^2 \times [-1, 1]$  of thickness $2$ with horizontal coordinates $x,y$ and vertical coordinate $z$.

Plane parallel shear flows are solutions of the non-dimensional stationary Navier-Stokes equations
\begin{eqnarray}\label{Couette-gen}
	\left\{ \begin{array}{l}
		{\bf U}\!\cdot\!\nabla {\bf U} ={\rm Re}^{-1} \Delta {\bf U}-\nabla P\\[5pt]
		\nabla \cdot {\bf U}=0 ,\\
	\end{array}  \right.
\end{eqnarray}
characterized by the functional form
\begin{align}
	\label{basic}
	{\bf U}= (	f(z),0,0) =  f(z) {\bf i},
\end{align}
where ${\bf U}$ is the velocity field and $P$ the pressure field, and ${\rm Re}$ is the Reynolds number. 
The function $f(z) : [-1, 1] \to \mathbb R$ is assumed to be sufficiently smooth and is called the shear profile. All the variables are written in a non-dimensional form. To non-dimensionalize the equations and the gap of the layer we use a Reynolds number based on the average shear and half gap $d$ (see \cite{FalsaperlaGiacobbeMulone2019}).

In particular, for fixed velocity at the boundaries $z=\pm 1$, there are two well known profiles:

a) Couette  $f(z)=z$,

b) Poiseuille $f (z) = 1-z^2$.	

\subsection{Perturbation equations}
The perturbation equations to the plane parallel shear flows, in non-dimensional form, are 
\begin{align}
	\label{Couette-gen}
	\left\{ \begin{array}{l}
		u_t = -  {\bf u}\!\cdot\!\nabla u+ {\rm Re}^{-1} \Delta u -  (f u_x+f' w)- p_x\\[5pt]
		v_t = -  {\bf u}\!\cdot\!\nabla v+ {\rm Re}^{-1} \Delta v -  f v_x - p_y\\[5pt]
		w_t = -  {\bf u}\!\cdot\!\nabla w+  {\rm Re}^{-1}\Delta w -  f w_x- p_z\\	[5pt]
		u_x+v_y+w_z=0 .\\	
	\end{array}  \right.
\end{align}
In \eqref{Couette-gen} ${\bf u}$ is  the perturbation velocity field.  It   has components  $(u, v, w)$ in the directions $x,y,z$, respectively. $p$ denotes the perturbation to the pressure field.

Here we use the symbols $g_x$ as $\frac {\partial g} {\partial x}$, $g_t$ as $\frac {\partial g} {\partial t}$, etc., for any function $g$.

To system (\ref{Couette-gen}) we append the \textit{rigid} boundary conditions $${\bf u}(x,y,\pm 1,t)=0, \q (x,y,t) \in  \R^2 \times (0, +\infty), $$ and the initial condition
$${\bf u}(x,y,z,0)= {\bf u_0}(x,y,z), \q {\rm in}  \q\mathcal D,$$
with ${\bf u_0}(x,y,z)$ solenoidal vector which vanishes at the boundaries.



Assume that both ${\bf u}$ and $\nabla p $ are $x,y$-periodic with periods $2\pi/a$ and $2\pi/b$ in the $x$ and $y$ directions, respectively,  with wave numbers $( a, b) \in \R^2_+$  . In the following it suffices therefore to consider functions over the periodicity cell 
$$\Omega= [0, \frac{2\pi}{a}]\times [0, \frac{2\pi}{ b}] \times [- 1, 1] .$$
As the basic function space, we take $L_2(\Omega)$, which is the space of square-summable functions in $\Omega$ with the scalar product denoted by
$$(g,h) = \int_0^{\frac{2\pi}{a}} \int_0^{\frac{2\pi}{ b}} \int_{-1}^1 g(x,y,z) h(x,y,z) dxdydz, $$ 
and the corresponding  norm $\Vert g \Vert = (g,g)^{1/2}.$

We recall the definitions of streamwise and spanwise perturbations.

\begin{defn} {\rm We define \emph{streamwise} (or longitudinal) perturbations the perturbations 	${\bf u}, p$ which do not depend on $x$.} 
\end{defn}

\begin{defn} {\rm We define \emph{spanwise} (or transverse) perturbations the perturbations 	${\bf u}, p$ which do not depend on $y$.  The \textit{2D spanwise} perturbations are the two-dimensional spanwise perturbations for which $v=0$.}
\end{defn}

Linear stability/instability is obtained by studying the linearized system neglecting the nonlinear terms in \eqref{Couette-gen}.

We recall that the classical results of \cite{Romanov1973}  prove that Couette flow is \textit{linearly stable} for \textit{any Reynolds} number, while Poiseuille flow is unstable for any Reynolds number bigger that $5772$, \cite{Orszag1971}. The last result is obtained for $2D$ spanwise perturbations.

The Squire theorem, \cite{Squire1933}, holds for the linearized system: the most destabilizing perturbations are two-dimensional spanwise  perturbations. 
Critical Reynolds numbers for Poiseuille flows can be obtained by solving the Orr-Sommerfeld equation, see Drazin and Reid \cite[p. 156]{DrazinReid2004}.

\section{Nonlinear monotone energy stability}
Here we  study \textit{ the  nonlinear monotone  energy stability} with the Lyapunov second method, by using the \textbf{energy} 
\be \label{EN-EQ}
E(t) = \dfrac{1}{2}[\Vert u \Vert^2 +  \Vert v \Vert^2 + \Vert w \Vert^2 ]. 
\ee
We obtain \textit{sufficient conditions of nonlinear monotone exponential stability}.

Taking into account the solenoidality of ${\bf u}$ and the boundary condition, we  write the Reynolds-Orr energy identity, \cite{Reynolds1895}
\begin{align}
	\label{Energy}
	\dot E= -(f'w,u) - {\rm Re}^{-1} [\Vert \nabla u \Vert^2+\Vert \nabla v \Vert^2+\Vert \nabla w \Vert^2]. 
\end{align}
For perturbations with  $ -(f'w,u) \le 0$ and $\Vert \nabla {\bf u} \Vert>0$, then $\dot E < 0$.

If $\Vert \nabla {\bf u} \Vert>0$, from \eqref{Energy}, we easily have

	\begin{eqnarray}
	\begin{array}{l}\label{ineq-}
		\dot E=  \left(\dfrac{-(f'w,u)}{\Vert \nabla u \Vert^2+\Vert \nabla v \Vert^2+\Vert \nabla w \Vert^2} - \dfrac{1}{{\rm Re}}\right)\Vert \nabla {\bf u} \Vert^2  \le \\[3mm]
		\le\left(m - \dfrac{1}{{\rm Re}}\right)\Vert \nabla {\bf u} \Vert^2 ,
	\end{array}
\end{eqnarray}
where
\begin{eqnarray} \label{maxprima}
	\dfrac{1}{{\rm Re}_E} = m= \max_{\cal S} \dfrac{-(f'w,u)}{\Vert \nabla u \Vert^2+\Vert \nabla v\Vert^2+\Vert \nabla w \Vert^2}, 
\end{eqnarray}
$\cal S$ is the space of the   {\textit{kinematically admissible fields}}  
\begin{eqnarray} \nonumber
	\begin{array}{l}
		\label{spaceS}
		{\cal S}= \{u, v, w \in H^1(\Omega), \; u=v=w=0 \q \hbox{on the boundaries,}\\[3mm] \hbox{ periodic in \textit{x}, and \textit{y},}  \q u_x+v_y+w_z=0,\q  \Vert \nabla {\bf u} \Vert>0\},
	\end{array}
\end{eqnarray}
with $H^1(\Omega)=W^{1,2}(\Omega)$ the usual Sobolev space: the subset of functions ${\displaystyle h} \in {\displaystyle L^{2}(\Omega )}$ such that ${\displaystyle h}$ and its weak derivatives up to order ${\displaystyle 1}$ have a finite $L^2$ norm.

We also observe that, taking into account the observations of Lorentz \cite{Lorentz.1907} and Lamb \cite{Lamb.1924}, the functions (u,v,w) in ${\cal S}$ must satisfy the scale invariance property.

Now, consider the functional ratio	
 \[\label{p1} 
	{\cal F}(u,v,w)=	\small \frac{-(f'w,u)}{\Vert u_x \Vert^2+\Vert u_y \Vert^2+\Vert u_z \Vert^2+\Vert v_x \Vert^2+\Vert v_y \Vert^2+\Vert v_z \Vert^2+\Vert w_x \Vert^2+\Vert w_y \Vert^2+\Vert w_z \Vert^2}  \]
	
	in ${\cal S}$.  
	
	We have:
	
	\begin{prop}
		There exists the maximum of ${\cal F}(u,v,w)$ in $\cal S$ (Rionero (1968) \cite{Rionero1968}) and the maximum is a non-negative value.
	\end{prop}
	
	Note that, by assuming  $u_z(z)\not=0$,  ${\cal F}(u(z),0,0)=0$.
	\begin{prop}
		The maximum of ${\cal F}(u,v,w)$ in $\cal S$ is  equal to the maximum of the ratio ${\cal F}(u,v,w)$ where now  $v_y=0$. 
		\end{prop}
	In fact, consider the denominator of ${\cal F}$
	
	$$\Vert u_x \Vert^2+\Vert u_y \Vert^2+\Vert u_z \Vert^2+\Vert v_x \Vert^2+\Vert v_y \Vert^2+\Vert v_z \Vert^2+\Vert w_x \Vert^2+\Vert w_y \Vert^2+\Vert w_z \Vert^2 ,$$ by substituting $v_y=-(u_x+w_z)$, we have
	
	$$\Vert u_x \Vert^2+\Vert u_y \Vert^2+\Vert u_z \Vert^2+\Vert v_x \Vert^2+\Vert u_x+w_z \Vert^2+\Vert v_z \Vert^2+\Vert w_x \Vert^2+\Vert w_y \Vert^2+\Vert w_z \Vert^2 .$$
For any $(u,v,w) \in {\cal S}$ we have

	\begin{eqnarray}
	\begin{array}{l}\label{ineq-00}
	\Vert u_x \Vert^2+\Vert u_y \Vert^2+\Vert u_z \Vert^2+\Vert v_x \Vert^2+\Vert u_x+w_z \Vert^2+\Vert v_z \Vert^2+\Vert w_x \Vert^2+\Vert w_y \Vert^2+\Vert w_z \Vert^2 	 \ge \\[3mm]
		\ge\Vert u_x \Vert^2+\Vert u_y \Vert^2+\Vert u_z \Vert^2+\Vert v_x \Vert^2+\Vert v_z \Vert^2+\Vert w_x \Vert^2+\Vert w_y \Vert^2+\Vert w_z \Vert^2  .
	\end{array}
\end{eqnarray}
Then supposing $-(f'u,w)\ge 0$, the maximum of ${\cal F}$ is less than or equal to the maximum of the ratio with  a field $(u,v,w)\in {\cal S}$ such that $v_y= -(u_x+w_z)=0$.

To prove this,  define 
 \[\label{p1} 
{\cal F}_x(u,v,w)=	\small \frac{-(f'w,u)}{\Vert u_x \Vert^2+\Vert u_y \Vert^2+\Vert u_z \Vert^2+\Vert v_x \Vert^2+\Vert v_z \Vert^2+\Vert w_x \Vert^2+\Vert w_y \Vert^2+\Vert w_z \Vert^2}  \]
ad observe that for \textit{any} fixed $(u_1,v_1,w_1) \in {\cal S}$ we have
$${\cal F}(u_1,v_1,w_1) \le {\cal F}_x(u_1,v_1,w_1). $$
Moreover, 
$${\cal F}_x(u_1,v_1,w_1) \le \max {\cal F}_x(u,v,w),$$
where the maximum is sought among all fields $(u,v,w) \in {\cal S}$ with $v_y=0$. Suppose that this maximum is obtained in $(\hat u,\hat v,\hat w)\in {\cal S}$ with $\hat v_y=0$. We therefore have
$${\cal F}(u_1,v_1,w_1) \le {\cal F}_x(\hat u,\hat v,\hat w), \quad \hbox{for \textit{any}} \quad (u_1,v_1,w_1) \in {\cal S}.$$

From this it follows that ${\cal F}_x(\hat u,\hat v,\hat w)$ is an upper bound for the numerical set described by ${\cal F}(u_1,v_1,w_1)$ as $(u_1,v_1,w_1)$ varies in ${\cal S}$. Consequently, for the maximum of ${\cal F} (u_1,v_1,w_1)$ in ${\cal S}$, which is the \textit{least} upper bound (and maximum), we have 
 $$ \max_{\cal S} {\cal F}(u_1,v_1,w_1) \le \max {\cal F}_x(u,v,w)={\cal F}_x(\hat u,\hat v,\hat w) .$$ Obviously, in the previous inequality the \textit{equal} sign holds because the set of elements $(u,v,w)$ of ${\cal S}$ with $v_y=0$ is a subset of ${\cal S}$.

\begin{prop}
	The maximum of ${\cal F}_x(u,v,w)$ in $\cal S$ is obtained at a vector field $(u(x,z), 0, w(x,z))$. It coincides with the maximum of ${\cal F}(u,v,w)$ assumed at two-dimensional spanwise perturbations, as Orr has supposed.
\end{prop}
First we prove that the maximum of  ${\cal F}_x(u,v,w)$  is obtained at a vector $(u(x,y,z), 0, w(x,y,z))$ of ${\cal S}$ such that $u_x+w_z=0$. To see this, we write the Euler-Lagrange equations of this maximum (let us call this maximum with the same symbol $m$ as before, in fact we will prove that it coincides with the maximum \eqref{maxprima})

\begin{eqnarray}  
	\dfrac{1}{{\rm Re}_E} = m= \max_{\cal S} \dfrac{-(f'w,u)}{\Vert \nabla u \Vert^2+\Vert v_x \Vert^2+ \Vert v_z \Vert^2+\Vert \nabla w \Vert^2}. 
\end{eqnarray}
They are
\begin{align}
	\label{EL-Orr0-0}
	(-f' w + 2m \Delta u) \, {\bf i}+ 2m (v_{xx}+v_{zz}) \,  {\bf j} + (-f' u  +2m \Delta w) \, {\bf k}  = \nabla \lambda,
\end{align}
where $\lambda (x,y,z)$ is a Lagrange multiplier.  

Suppose that $(\hat{u}, \hat{v}, \hat{w})$ is a maximizing vector field, with $\hat{v}_y=-(\hat{u}_x+\hat{w}_z)=0$ and let  $m={\cal F}_x(\hat{u}, \hat{v}, \hat{w})$, 
\be\label{max-cal-0}
(f'\hat{w},\hat{u}) +m [ \Vert \nabla \hat{u}\Vert^2 + \Vert \nabla \hat{w}\Vert^2]=0, 
\ee
and 
\be 
2m [\Vert \hat{v}_{x}\Vert^2+\Vert \hat{v}_{z}\Vert^2]= (\lambda_y, \hat v)= - (\lambda,\hat v_y)=0.
\ee
Thus $\hat v=0$.  From
\eqref{EL-Orr0-0}, we have
\begin{align}
	\label{elwz-comp}
	\begin{cases}
		-f'\hat{w}+2 m \Delta \hat{u}&= \lambda_x\\
		-f'\hat{u}  + 2 m \Delta \hat{w} &= \lambda_z .
	\end{cases}
\end{align}

We adopt plane-form perturbations
$g(x,y,z)= G(z) \exp\{ ia x+ib y\}$, with $a$ and $b$ wave numbers in the $x$ and $y$ directions and $g(x,y,z)$ is any function $u, v, w, \lambda$. 

Suppressing the \textit{hats}, assuming that $ u, v, w $ could depend on all the variables $ x, y, z $, taking the third component of the curl and the third component of the double curl, to eliminate the multiplier, we obtain the system (for simplicity we refer to the \textit{Couette flow}, $ f '= 1 $, but the proof does not change for Poiseuille flow)
\begin{align}
	\label{elwz-55}
	\begin{cases}
		w_y-2m \Delta u_y   =0\\
		\Delta_1 u  -2 m \Delta \Delta_1 w -w_{xz}+ 2m \Delta u_{xz}=0 ,
	\end{cases}
\end{align}
with the constraint $ u_x +  w_z = 0 $, and $ \Delta_1 $ is the two-dimensional Laplacian in $ x $ and $ y $.

From the first equation we have:

i) $ b = 0 $, which implies that  $ \Delta_1 u = u_ {xx} = - w_ {xz} $, $ \Delta u_ {xz} = - \Delta w_ {zz} $, and the resulting equation is the Orr's equation

$$ m \Delta \Delta w + w_ {xz} = 0. $$

ii) If $ b \not= 0 $ then $ w-2m \Delta u = 0 $, and system \eqref{elwz-55} becomes 
\begin{align}
	\label{elwz-56}
	\begin{cases}
		w-2m \Delta u   =0\\
		\Delta_1 u  -2 m \Delta \Delta_1 w -w_{xz}+ 2m \Delta u_{xz}=0 ,
	\end{cases}
\end{align}
with the solenoidality condition $u_x+w_z=0$.
Obviously if $m=0$ the functional ratio is zero. 

Otherwise ($m\not=0$), taking into account the first equation, the second becomes $\Delta_1 u  -2 m \Delta \Delta_1 w -w_{xz}+ 2m \Delta u_{xz}= \Delta_1 (u-2m \Delta w) - (w-2m \Delta u)_{xz}= \Delta_1 (u-2m \Delta w)=0$. Therefore we have the system 
\begin{align}
	\label{elwz-56}
	\begin{cases}
		w-2m \Delta u   =0\\
	u-2m \Delta w=0 \\
	u_x+w_z=0.
	\end{cases}
\end{align}
From these three equations and the boundary conditions $u=w=w_z=0$ on the boundaries $z=\pm 1$, it follows that \textit{all the derivatives with respect to $z$} of $w$ are zero on the boundaries. For instance, by calculating the equation $u-2m \Delta w=0$ on the boundaries it follows that $w''=0$. Taking the partial derivative with respect to $x$ in the first equation we have $w_x-2m \Delta u_x=0$. Since $u_x=-w_z$, we have $w_x+2m \Delta w_z=0$. From this we have $w'''=0$ on the boundaries. By taking further derivatives with respect to $z$, we obtain that all the derivatives of $w$ are zero on the boundaries.   Therefore, $w = 0$ and $ u = 0 $ in $\Omega$. This implies that $\Vert \nabla {\bf u}\Vert=0$, that has been excluded. Therefore, the derivatives of $u$ and $w$ with respect to $y$ must be zero.  Hence, we obtain the same critical Reynolds number of Orr on  2-dimensional spanwise perturbations. 

Note the maximum of ${\cal F}$ \textit{on the subspace ${\cal S}_0$} of ${\cal S}$ of function of ${\cal S}$ such that $(u(x,z),0,w(x,z))\in {\cal S}$, $u_x+w_z=0$ is \textit{exactly} $1/ {\rm Re}_E$.

\begin{prop}
	The critical Reynolds number ${\rm Re}_E$ is given by the Orr's equation 
\begin{align} 
	\label{Orr-eq}
	{\rm Re}_E (f'' w_x+2f' w_{xz})+ 2\Delta \Delta w=0.	
\end{align}	with b.c. $w=w'=0$.
\end{prop}
Finally, we have the proposition:
\begin{prop}
	On the streamwise perturbations (those which satisfy the condition $\frac{\partial}{\partial x}\equiv 0)$ the scalar product $-(f' u,w) \le 0$. 
\end{prop}
See \cite{FalsaperlaMulonePerrone2022}, Theorem 3.1.
\vskip .2cm
From inequality \eqref{ineq-} and the Poincaré's inequality we obtain
\begin{align}
	\label{time-en11}
	\begin{array}{l}
		\dot E	\le  \dfrac{\pi^2}{4}\left( \dfrac{1}{{\rm Re}_E} - \dfrac{1}{{\rm Re}}\right)\left[\Vert u \Vert^2 + \Vert v \Vert^2+\Vert w \Vert^2\right]= \dfrac{\pi^2}{2} \left( \dfrac{1}{{\rm Re}_E} - \dfrac{1}{{\rm Re}}\right) E  .
	\end{array}
\end{align} 
By integrating this inequality, we get:
\begin{thm}\label{t31}
	\textit{Assuming ${\rm Re} <{{\rm Re}_E}$, the  basic shear flow (\ref{basic}) is nonlinear monotone exponentially stable according to the classical energy:
		\be\label{mon-stab} E(t) \le E(0) \exp\left\{{\dfrac{\pi^2}{2}\left( \dfrac{1}{{\rm Re}_E} - \dfrac{1}{{\rm Re}}\right)t}\right\}, \q \forall t \ge 0.
		\ee 
} \end{thm}

Hence, the critical nonlinear  monotone energy Reynolds number ${\rm Re}_E$ is the Orr critical value ${\rm Re_{Orr}}$,  and  ${\rm Re_{Orr}}= 44.3$ for Couette flow and ${\rm Re_{Orr}}= 87.6$ for Poiseuille flow.  Moreover, from \eqref{time-en} we have the following Corollary.

We note that, if ${\rm Re} <{{\rm Re}_E}$, it is not possible to have  transient maximum growth of $E(t)$ greater than $1$. However, this is not excluded for ${\rm Re} >{{\rm Re}_E}$, see \cite{ReddyHenningson1993}.
\begin{corollary}
	\textbf{(Squire Theorem} for nonlinear energy stability). 
	
	\textit{The least stabilizing perturbations in the energy norm \eqref{EN-EQ} are two-dimensional perturbations  which do not depend on $y$, the spanwise  perturbations.
	}
\end{corollary}

\subsection{Partition of dissipative term}

Here we  give a \textit{sufficient condition of nonlinear monotone energy stability } of the base Couette flow and Poiseuille flows. We introduce  a partition of  the dissipation term into two terms. For this, we use  the Reynolds-Orr energy identity, \cite{Reynolds1895}
\begin{align}
	\label{Energy1}
	\dot E= -(f'w,u) - {\rm Re}^{-1} [\Vert \nabla u \Vert^2+\Vert \nabla v \Vert^2+\Vert \nabla w \Vert^2]. 
\end{align}
If $ -(f'w,u) \le 0$ and $\Vert \nabla {\bf u} \Vert>0$, then $\dot E < 0$. In the case
$ -(f'w,u) > 0$ we \textit{split the dissipation term} in the energy identity in this way:
\begin{align}
	\label{Energy-split}
	\dot E= -(f'w,u) - {\rm Re}^{-1} [\Vert \nabla u \Vert^2+\Vert \nabla w \Vert^2] -{\rm Re}^{-1}\Vert \nabla v \Vert^2. 
\end{align}
We have 
\begin{align}
	\label{time-en}
	\begin{array}{l}
		\dot E= -(f'w,u) - {\rm Re}^{-1} [\Vert \nabla u \Vert^2+\Vert \nabla w \Vert^2] -{\rm Re}^{-1}\Vert \nabla v \Vert^2 =\\[3mm]
		= \left(\dfrac{-(f'w,u)}{\Vert \nabla u \Vert^2+\Vert \nabla w \Vert^2} - \dfrac{1}{{\rm Re}}\right)[\Vert \nabla u \Vert^2 + \Vert \nabla w \Vert^2] - {\rm Re}^{-1}\Vert \nabla v \Vert^2\le \\[3mm]
		\le \left( \dfrac{1}{{\rm Re}_1} - \dfrac{1}{{\rm Re}}\right)[\Vert \nabla u \Vert^2 + \Vert \nabla w \Vert^2] - {\rm Re}^{-1}\Vert \nabla v \Vert^2 ,
	\end{array}
\end{align} 
where 
\begin{align}
	\label{maxRefin}
	\begin{array}{l}
		m_1=\dfrac{1}{{\rm Re}_1}= 
		\max_{\cal S}	
		\dfrac{-(f'w,u)}{\Vert u_x \Vert^2+\Vert u_y \Vert^2 + \Vert u_z \Vert^2 + \Vert w_x \Vert^2+\Vert w_y \Vert^2	+\Vert w_z \Vert^2},
	\end{array}
\end{align}
and ${\cal S}$ is the space of the kinematically admissible fields introduced above.

Now we use the divergence-free condition $u_x+v_y+w_z=0$, and write the maximum problem \eqref{maxRefin} in the following way
	\begin{align}
			\label{maxRefin-0}
			\begin{array}{l}
					m_1=\dfrac{1}{{\rm Re}_1}= 
					\max_{\cal S}	
					\dfrac{-(f'w,u)}{\Vert u_x \Vert^2+\Vert u_y \Vert^2 + \Vert u_z \Vert^2 + \Vert w_x \Vert^2+\Vert w_y \Vert^2	+\Vert w_z \Vert^2}=\\[15pt]
					=\max_{\cal S}	
					\dfrac{-(f'w,u)}{\Vert v_y \Vert^2+2(v_y,w_z)+2\Vert w_z \Vert^2 + \Vert u_y \Vert^2 + \Vert u_z \Vert^2+\Vert w_x \Vert^2	+\Vert w_y \Vert^2}.
				\end{array}
		\end{align}
	Applying the Cauchy-Schwarz and the Poincar\'{e} inequalities to the numerator of the functional ratio \eqref{maxRefin-0}  we can prove that it has a finite upper bound and that the maximum exists, see \cite{Rionero1968},  \cite{GaldiRionero1985}.

	The Euler-Lagrange equations of this maximum problem are
	\begin{align}
			\label{elwz-4}
			\begin{cases}
					-f' w+2m_1 (u_{yy}+u_{zz})   =\lambda_x\\
					2 m_1 (w_{zy}+v_{yy}) =\lambda_y\\
					-	f' u  +2 m_1 v_{yz}+4m_1 w_{zz}+2m_1 (w_{xx}+w_{yy})=\lambda_z .
				\end{cases}
		\end{align}
	From the second equation and the solenoidality of ${\bf u}$,  we have $(\lambda+2m_1 u_x)_y=0$. This gives $\lambda= -2m_1 u_x + h(x,z)$ with $h(x,z)$ an \textit{arbitrary} function of $(x,z)$.
	
	Substituting $\lambda$ in \eqref{elwz-4}, we easily obtain
	\begin{align}
			\label{elwz-5}
			\begin{cases}
					-f' w+2m_1 \Delta u   =h_x\\
					-	f' u  +2 m_1 \Delta w=h_z .
				\end{cases}
		\end{align}
	These equations coincide with \eqref{elwz-comp}$_{1,3}$. Moreover, we now prove that $u_x+w_z=-v_y=0$. In fact, form \eqref{elwz-5}, we have
\begin{align}
	\label{elwz-6}
	\begin{cases}
		(-f' w,u)=2m_1 \vert \nabla u \vert^2  +(h_x,u)\\
		(-f' w,u)=2m_1  \vert \nabla w \vert^2  +(h_z,w) .
	\end{cases}
\end{align}
By summing these equations, and taking into account that $m_1$ is the maximum, we easily obtain

\[
0= -(f' u,w)-m_1 (\Vert \nabla u \Vert^2+ \Vert \nabla w \Vert^2 )= - \frac{1}{2} (h, u_x+w_z).
\]
This implies that $u_x+w_z=0$ (note that $h$ is an arbitrary function of $(x,z))$.

If $h\not= 0$, for this linear system we can now use, as in the proof of Proposition 3.3, the representation $ g (x, y, z) = G (z) \exp\{ia x + ib y \} $, for $u,w,h$,  with $ a $ and $ b $ wave numbers, and  where now $b=0$. 
This implies that $u$ and $w$ do not depend on $y$, and the resulting equation is that of Orr

$$ m_1 \Delta \Delta w + w_{xz} = 0, $$
where $\Delta w= w_{xx}+ w_{zz}$.

If $ h= 0, $ we obtain  system \eqref{elwz-56} (in the general case: Couette or Poiseuille). We can then proceed step by step with the proof in Proposition 3.3 above from system \eqref{elwz-56} to the end of the proof of Proposition 3.3 (with some more calculations in the case of Poiseuille flow).

Therefore,  the maximum  $ m_1$ is the value computed by Orr and it is equal to $m$ in Proposition 3.3.

\begin{figure}{\label{tilted2}}
\begin{center}
\includegraphics[width=10cm]{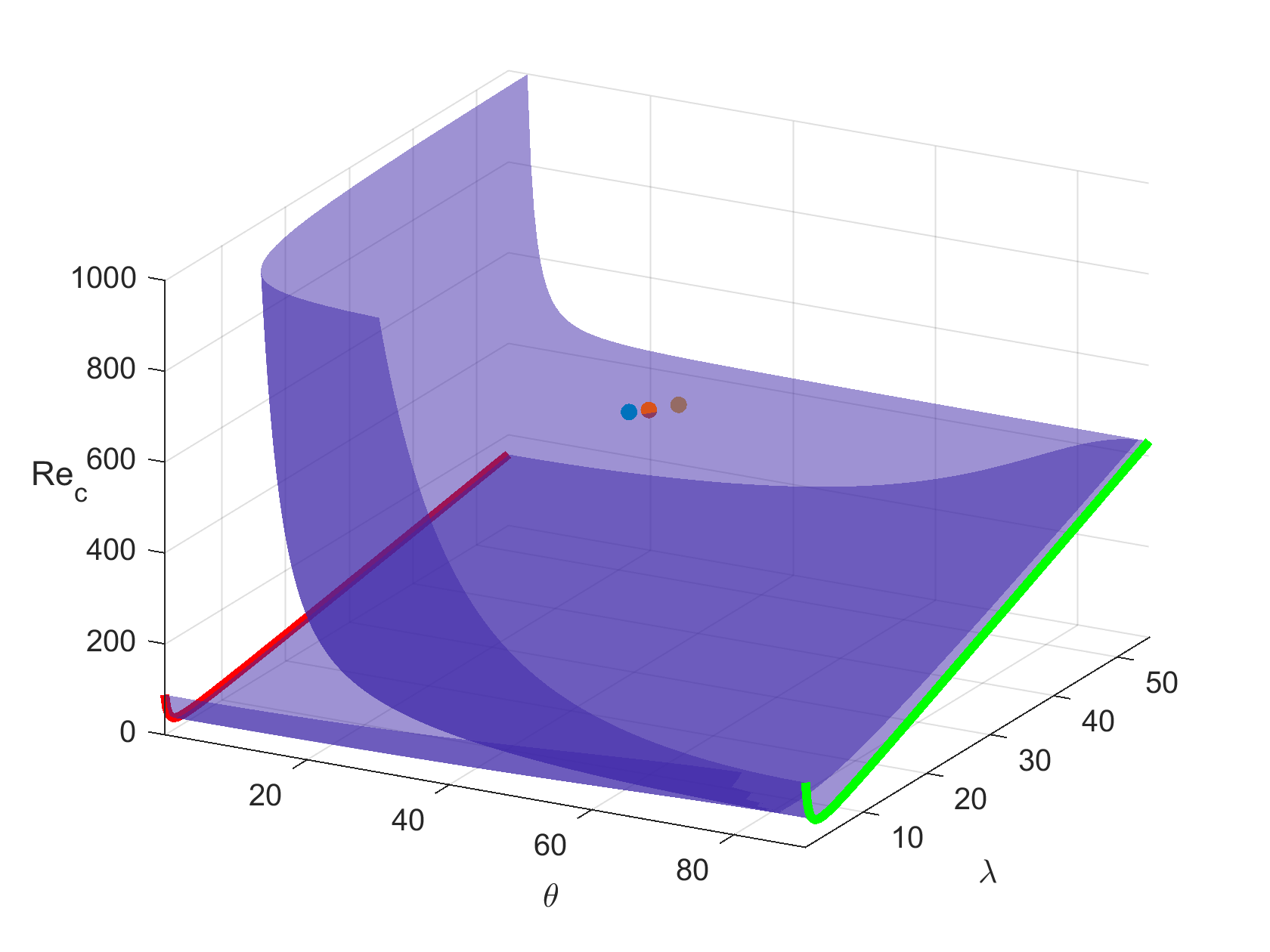}
\end{center}  \caption{Plane Couette critical energy Orr-Reynolds  numbers ${\rm Re}=\rm Re_c$ as a function of the wavelength $\bar \lambda$ and angle $\theta$. 
The three points (blue, orange and brown)  refer to the cases $\theta=25^\circ$, $\bar \lambda= 46$, and $\theta=26^\circ$, $\bar \lambda= 48$, $\theta=27,5^\circ$, $\bar \lambda= 51$. The corresponding experimental Reynolds numbers are 395, 385, 375, respectively.}
\end{figure}

\section{Conclusion}

We have studied the problem of nonlinear monotone energy  stability of laminar flows of Couette and Poiseuille. We have proved in two different ways that the  critical Reynolds number for nonlinear stability is the one obtained by Orr on two-dimensional spanwise perturbations, as also Serrin has probably thought.

This conclusion contradicts both Joseph's result and, in part, the results obtained by many authors in recent years on the initial energy growth for Reynolds number values between the critical values of Joseph and Orr.

We note that in the literature reference is made to Busse's rigorous result to validate Joseph's result. However, we observe that Busse studies the minimum problem for $ Re $ taking the production term $ - (f'u, w) $ in absolute value.
In this way he neglects the classes of perturbations for which the numerator is less than or equal to zero. These include streamwise perturbations, as we have shown. This probably led to Joseph's erroneous result and the gap in his proof that Serrin talks about.

It should also be noted that if we had considered the maximum of the ratio
$$\dfrac{(f'w,u)}{\Vert \nabla u \Vert^2+\Vert \nabla v\Vert^2+\Vert \nabla w \Vert^2} ,$$
 (note that the sign in the numerator has changed), we would have obtained that on the solutions of the Euler-Lagrange equations of the streamwise type (considered by Joseph) we would have obtained that the maximum is the same as that of the functional $ {\cal F} $. Therefore, for these streamwise functions, $ m $ is the maximum of the functional obtained from $ {\cal F} $ by replacing $ - (f' u, w) $ with $ \vert -(f' u, w) \vert $ which is exactly the maximum  studied by Busse \cite{Busse1972} (see also \cite{ReddyHenningson1993}).
 
 \vskip .3cm

Here we have performed a proof which indicates nonlinear monotone  energy stability for every Reynolds number below Orr's threshold. Other than being a mathematical argument, this fact can be compared with the instability thresholds obtained experimentally in \cite{Prigent.et.al2003}. In \cite{FalsaperlaGiacobbeMulone2019} the authors have proposed a formula that, using the critical values obtained with Orr's approach, generates a Reynolds instability surface depending on inclination and width of the instability rolls on which the experimental data fit excellently, while they do not fit in the classical Reynolds instability surface coming from a direct analysis of the orbital derivative of the energy function (see Figure 1).  Fig. 1 shows two surfaces; the one below represents in polar coordinates the critical surface $R_1 (\alpha, \beta)$ described in \cite[pag. 213]{ReddyHenningson1993} and in \cite[Fig. 1]{FalsaperlaMulonePerrone2022}.
The surface above is obtained from the formula given in \cite{FalsaperlaGiacobbeMulone2019}  for a tilted perturbation of an angle  
$\theta= arcsin( b/ \bar \lambda)$ with $\bar \lambda = \sqrt{a^2+b^2}$, $a$ and $b$ wave-numbers in the $x$ and $y$ directions.

\begin{eqnarray} \nonumber
		{\rm Re}_c={\rm Re}_{Orr}\left(\frac{2\pi}{\bar \lambda\sin \theta}\right)/\sin \theta.
	\end{eqnarray}

The surfaces shown in Figure 1 give the critical Reynolds numbers obtained by Joseph (the lower one) and by Orr (the upper one).
The three points corresponding to given angles and wavelengths of the experiments of \cite{Prigent.et.al2003} are very close to the upper surface (the Orr's surface). 

\hskip .2cm

The results I have obtained in the present paper are in contrast with Joseph's calculations and the growth of energy  on the streamwise perturbations. The interpretation of these facts needs a deeper investigation, in particular on the role that nonlinear terms play in the perturbation equations.  I plan to perform this in further works.

I also observe that a corollary of our result is the validity of  Squire theorem for nonlinear system. 

Finally, I note that similar results hold: for shear flows in a channel  with different boundary conditions (for instance with stress-free b.c.), and for shear flows in magneto-hydrodynamics (forthcoming paper).

\vskip .4cm

\textbf{Statements and Declarations}

We declare that we have no competing financial and / or non-financial interests.
\vskip.4cm

\textit{Data sharing not applicable to this article as no datasets were generated or analysed during the current study.}

\vskip .4cm
{\bf Acknowledgments}\\
{I thank two anonymous referees for their comments and
	highly appreciate their suggestions, which significantly contributed
	to improving the quality of the publication.
	
	The research that led to the present paper was partially supported by the following Grants: 2017YBKNCE of national project PRIN of Italian Ministry for University and Research, PTR DMI-53722122146 "ASDeA" of the University of Catania. We also thank the group GNFM of INdAM for financial support.
}


\begin{thebibliography}{99}

\expandafter\ifx\csname natexlab\endcsname\relax
\def\natexlab#1{#1}\fi
\expandafter\ifx\csname selectlanguage\endcsname\relax
\def\selectlanguage#1{\relax}\fi

\bibitem{Busse1972} {F.H. Busse} (1972)  \textit{A Property of the energy stability limit for plane parallel shear flow}, {Arch. Rat. Mech. Anal.}, {47} (1),  pp. 28--35.

\bibitem{ButlerFarrel1992} {M. Butler  and B. F. Farrel } (1992)  \textit{Three-dimensional optimal perturbations in viscous shear flow}, { Phys. Fluids} {4}, p. 1637 .

\bibitem{Chandrasekhar1961} {S. Chandrasekhar }  \textit{Hydrodynamic and hydromagnetic stability}, Oxford, Clarendon Press: Oxford University Press (1961).

\bibitem{DrazinReid2004} {P.G. Drazin  and W.H. Reid} {\it Hydrodynamic Stability}, Cambridge Monographs on Mechanics, Cambridge University Press, 2nd Ed. (2004) .

\bibitem{Eckhardt.et.al2007} {B. Eckhardt ,  T. M. Schneider, B. Hof and  J. Westerweel } (2007)  \textit{Turbulence Transition in Pipe Flow}, { Annu. Rev. Fluid Mech.} {39}  pp. 447--468. 

\bibitem{FalsaperlaGiacobbeMulone2019} {P. Falsaperla, A. Giacobbe  and G.  Mulone} (2019)\textit{ Nonlinear stability results for plane Couette and Poiseuille flows}, {Physical Review E}, {100},  013113.  https://doi.org/10.1103/PhysRevE.100.013113 

\bibitem{FalsaperlaMulonePerrone2022} {P. Falsaperla, G. Mulone and C. Perrone} 
(2022) \textit{Energy stability of plane Couette and Poiseuille flows: a conjecture}, {European J. Mech./B Fluids}, {93} pp. 93--100  https://doi.org/10.1016/j.euromechflu.2022.01.006

\bibitem{Fuentes.Goluskin.Chenishenko.2022} F. Fuentes, D. Goluskin and S. Chernyshenko, (2022) \textit{Global stability of fluid flows despite transient growth of energy}. Phys. Rev. Lett. 128, p. 204502.

\bibitem{GaldiRionero1985} {G.P. Galdi and S.  Rionero} {\it Weighted energy methods in 	fluid dynamics and elasticity}, Lecture Notes in Mat. vol. { 1134}, Springer-Verlag, New-York (1985) 

\bibitem{Joseph1966} { D.D. Joseph} (1966)  \textit{Eigenvalue bounds for the Orr-Sommerfeld equation}, {J. Fluid Mech.}, {33} part 3,  pp. 617--621. 

\bibitem{JosephCarmi1969} {D.D. Joseph and S. Carmi} (1969) \textit{Stability of Poiseuille flow in pipes, annuli and channels,} {Quart. App. Math.}, {26}, pp. 575--579.

\bibitem{Joseph1976} {D.D. Joseph}  {\it Stability of fluid motions}, vol. 1. Berlin, Germany: Springer (1976).

\bibitem{KaiserTilgnerVonWhal2005} {R. Kaiser, A. Tilgner and W.  vonWahl} (2005)  \textit{A generalized energy functional for plane Couette flow}, {SIAM J. Math. Anal.}, {37} (2), pp. 438--454 (doi:10.1137/S0036141004442604)

\bibitem{KaiserMulone2005} {R. Kaiser and G.  Mulone}  (2005) \textit{A note on nonlinear stability of plane parallel shear flows}, {J. Math. Anal. Appl.}, {302},  pp. 543--556.  https://doi.org/10.1016/j.jmaa.2004.08.025


\bibitem{Klotz.et.al2021} {L. Klotz,  A. M. Pavlenko  and J. E. Wesfreid} (2021) \textit{Experimental measurements in plane Couette-Poiseuille flow: dynamics of the large- and small-scale flow},  {J. Fluid Mech.}, {912},  pp. A-24-1--A-24-31   doi:10.1017/jfm.2020.1089.

\bibitem{Lamb.1924} H. Lamb \textit{Hydrodynamics}, fifth ed., Cambridge Univ. Press  (1924).

\bibitem{Lorentz.1907} H. Lorentz ( 1907) \textit{Ueber die Entstehung turbulenter Fl\"{u}ssigkeitschewegungen und \"{u}ber den Einfluss dieser Bewegungen bei der Str\"{o}mung durch R\"{o}hren}, Abhandlungen über theoretische Physik, Leipzig, 1907, i, 43.

\bibitem{Martinelli2011} {F. Martinelli, M. Quadrio, J.  McKernan and   . F. Whidborne} (2011)   \textit{Linear feedback control of transient energy growth and control performance limitations in subcritical plane Poiseuille flow}, { Phys. Fluids} {23}   pp. 014103-1 -- 014103-9.

\bibitem{Moffatt1990} {K. Moffatt}  \textit{Fixed points of turbulent dynamical systems and suppression of nonlinearity},  In {Whither turbulence}, J. Lumley (ed), Springer, 250 (1990).

\bibitem{Nagy.2022} P.T. Nagy (2022) \textit{Enstrophy change of the Reynolds-Orr solution in channel flow}. Phys. Rev. E 105, p. 035108.

\bibitem{Orr1907} {W. M'F. Orr} (1907) \textit{The stability or instability of the steady motions of a perfect liquid and of a viscous liquid}, {Proc. Roy. Irish Acad. A}, {27},  pp. 9--68,  pp. 69--138.

\bibitem{Orszag1971} {S.A. Orszag }  (1971) \textit{Accurate solution of the Orr-Sommerfeld stability equation}, {J. Fluid Mech.}, {50},  pp. 689--703. 

\bibitem{Prigent.et.al2003} {A. Prigent, G. Gr\'{e}goire, H. Chat\'{e} and O. Dauchot}  (2003) \textit{Longwavelength modulation of turbulent shear flows},  { Physica D}, {174}, pp. 100--113. DOI: 10.1016/S0167-2789(02)00685-1

\bibitem{ReddyHenningson1993} {S. C. Reddy and D. S.   Henningson}  (1993)
\textit{Energy growth in viscous channel flows}, {J. Fluid Mech.}, {252}, pp. 209--238.

\bibitem{Reynolds1895} {O. Reynolds}  (1895) \textit{On the dynamical theory of incompressible viscous fluids and the determination of the criterion}, Phil. Trans. Roy. Soc. A,  {186}, pp. 123--164. 

\bibitem{Rionero1968} {S. Rionero} (1968)   \textit{Metodi variazionali per la stabilit\`a asintotica in media in magnetoidrodinamica}, {Ann. Mat. Pura Appl}, {78}, pp. 339--364.

\bibitem{Romanov1973} {V. Romanov} (1973) \textit{Stability of plane-parallel Couette flow},  { Funct. Anal. Appl.}, {7},  pp. 137--146. 

\bibitem{SchmidHenningson2001} {P. J. Schmid and  D. S. Henningson }  
\textit{Stability and Transition in Shear Flows,  Applied Mathematical Sciences} (AMS, vol. {142}),  Springer-Verlag New-York, Inc. (2001).

\bibitem{Squire1933} {H.B. Squire}  (1933) \textit{On the stability of three-dimensional disturbances of viscous flow between parallel walls}, {Proc. Roy. Soc. A}, {142},  pp. 621--628. 

\bibitem{Tuckerman.et.al2020} {L. S. Tuckerman, M. Chantry and D. Barkley }  (2020) \textit{Patterns in Wall-Bounded Shear Flows},  { Annu. Rev. Fluid Mech.}, {52},  pp. 343--367.  

\bibitem{Xiong.Chen.2019} X. Xiong and  Z.M. Chen (2019) \textit{A conjecture on the least stable mode for the energy stability of plane parallel flows}. Journal of Fluid Mechanics 881, pp. 794-814. 


\end{thebibliography}
\end{document}